\documentclass[preprint,12pt]{aastex} 
\usepackage{color,epsfig} 
\definecolor{brown}{rgb}{0.6,0.4,0.2} 
\definecolor{purple}{rgb}{0.5,0,0.5} 
\shorttitle{Carbon Monoxide in Cas A} 
\shortauthors{Rho et al.}

\newcommand{\kms}{km~s$^{-1}$}

\newcommand{\spitzer}{\textit{Spitzer}} 
 
\newcommand{\mic}{$\mu$m} 
\shorttitle{Spectroscopic detection of Carbon Monoxide in Cas A }

\begin{document}

\title{Spectroscopic detection of Carbon Monoxide in the Young
  Supernova Remnant Cassiopeia A}

\author{ 
J. Rho\altaffilmark{1,4},  
T. Onaka\altaffilmark{2}, 
J. Cami\altaffilmark{3,4}
and W. T. Reach\altaffilmark{1}
} 
\altaffiltext{1}{Stratospheric Observatory for Infrared Astronomy, 
Universities Space Research Association, NASA Ames Research Center, MS 211-3,
Moffett Field, CA 94035; jrho, wreach@sofia.usra.edu} 
\altaffiltext{2}{Department of Astronomy, Graduate School of Science, The University of Tokyo, 7-3-1
Hongo, Bunkyo-ku, Tokyo 113-0033, Japan, onaka@astron.s.u-tokyo.ac.jp}
\altaffiltext{3}{Department of Physics \& Astronomy, University of
  Western Ontario, London, ON N6A 3K7, Canada ; jcami@uwo.ca }
\altaffiltext{4}{SETI Institute, 189 N. Bernardo Ave, Suite 100,
  Mountain View, CA 94043}

\begin{abstract} 

  We report the detection of carbon monoxide (CO) emission from the
  young supernova remnant Cassiopeia A (Cas A) at wavelengths
  corresponding to the fundamental vibrational mode at 4.65\,\mic.  We
  obtained {\it AKARI} Infrared Camera spectra towards 4 positions
  which unambiguously reveal the broad characteristic CO
  ro-vibrational band profile. The observed positions include
  unshocked ejecta at the center, indicating that CO molecules form in
  the ejecta at an early phase. We extracted a dozen spectra across
  Cas A along the long 1$'$ slits, and compared these to simple CO
  emission models in Local Thermodynamic Equilibrium to obtain
  first-order estimates of the excitation temperatures and CO masses
  involved. Our observations suggest that significant amounts of
  carbon may have been locked up in CO since the explosion 330 years
  ago. Surprisingly, CO has not been efficiently destroyed by
  reactions with ionized He or the energetic electrons created by the
  decay of the radiative nuclei. Our CO detection thus implies that
  less carbon is available to form carbonaceous dust in supernovae
  than is currently thought and that molecular gas
    could lock up a significant amount of heavy elements in supernova
    ejecta.

\end{abstract}

\keywords{molecular processes - ISM:molecules - supernova remnants:Cas A -- ISM:dust }

\section{Introduction}

Molecular emission from supernovae (SNe) remains rarely detected, as the
formation of even the most stable molecules is difficult in the SN ejecta.
The high UV radiation due to recombinations, and the energetic electrons
from the radioactive decay of $^{56}$Ni and $^{56}$Co dissociate molecules
efficiently. The presence of molecules in ejecta could thus imply an
inhomogeneous distribution of material within the ejecta and could
constrain the degree of mixing. 
Since CO is an effective coolant
\citep{liu95}, it enhances dust formation and thus the formation efficiency
of CO critically affects the dust formation efficiency.
Because of the rapid decrease in density and
  temperature during condensation in the ejecta, dust
  formation happens through non-equilibrium condensation. There are
  then two important key questions for dust condensation in
  supernovae. 
First, to determine which molecular species are relevant in
non-equilibrium condensation and second, whether the gas in the ejecta
is able to mix with the molecular species.


There are only about a dozen SNe known where CO has been detected. 
SN1987A is the only SN with spectroscopic detections of both the CO
fundamental and first overtone bands \citep{wooden93, spyromilio88}. Other
detections of the fundamental band have been made with the {\it Spitzer}
Space Telescope, but these observations cover only the tail of CO feature
at wavelengths longwards of 5.3\,\mic\ for SN2004dj, SN 2005aj, and
SN2006et \citep{kotak06}. More detections have been reported in the first
overtone band in the near-infrared, with observations of SN 1995ad, SN
1998S, SN 1998dl, SN 1999em, SN 1999gi, and SN 2000ew \citep{gerardy02}. 
In all cases, the CO detections correspond to times between 97- 200 days
after the explosion.

We previously detected CO in Cas A in near-infrared narrow-band images
centered at the 2.29\,\mic\ first overtone CO band \citep{rho09}. In this
paper, we report the spectroscopic detection of the fundamental CO band in
Cas A with the AKARI Infrared Camera. 

\section{Observations} 
We performed AKARI Infrared Camera (IRC; Onaka et al.~2007)
observations towards 4 positions in Cas A. Three positions are the
brightest knots in the near-infrared 2.29\,\mic\ narrow-band Palomar
image \citep{rho09}; two positions are towards the northern shell
(filenames are $``$n1" and $``$n2"), one towards the south ($``$s1"),
and an additional position towards the central position where
unshocked silicon ejecta (sipeak) are detected \citep{rho08}.

\begin{deluxetable}{lllllllll}
\tabletypesize{\scriptsize}
\rotate
\tablewidth{0pt}
\tablecaption{ AKARI observations and Properties of CO in Cas A}
\label{coproperties}
\tablehead{\colhead{filename} & \colhead{location} & \colhead{position}  & \colhead{Observation ID, Date$^a$ }          &  
\colhead{Velocity (km~s$^{-1}$)}   &
\colhead{T$^d$(CO)  }  & \colhead{CO mass$^{e}$} \\
    \colhead{}    & \colhead{}  &  \colhead{}   &  \colhead{} & \colhead{CO {\it (Ne)$^b$} (Ar) {\it [Si]$^c$}   }     & \colhead{(K)} & \colhead{(M$_\odot$)} \\
    }
\startdata
COn1hi  &N shell & 23:23:23.0, 58:50:27 & 3340001.1,01-17 &{\bf $^f$-1980 (-1697)(-1976)} {\it [-2116}, 2302]  &1500  &$3\times 10^{-10}$  \\
COn1hi2 & Nshell  &23:23:24.1, 58:50:35 &3340001.1,01-17 &{\bf -1990}  ({\bf -1687}) (-1863) [{\it -1700},{\it 2821}&1300&$4\times 10^{-10}$\\
COn2hi  &N inner-shell&23:23:22.4, 58:50:29 &3340001.2, 01-17   & -3810 {\it (-1512)} (-1527) {\it [-2330, 1777]}  &1300 &$7\times 10^{-10}$\\
COs1H0  & S inner   &23:23:28.1, 58:47:44 &3340005.1, 01-16    &  -4800 {\it (--)}$^f$  (--)$^f$ {\it [-907, 3442]}             &900  &$6\times 10^{-9}$\\
COs1H1 & S shell    &23:23:29.3, 58:47:14 &3340005.1, 01-16   &-3240 {\it (-1323, 1684)} (-1238) {\it [-770,  2439]}  & 1400 & $3\times 10^{-9}$ \\
\hline
COn1low &N shell      &23:23:23.1, 58:50:25 &3340002.1[2.2], 01-18  & {\bf  -5660} {\it (-1854)}{\bf (-1887, -7044) }{\it [--]$^g$} &2400  & $2\times 10^{-10}$\\
COn2low &N inner-shell&23:23:25.8, 58:50:05 &3340004.1, 07-21 & -5410 {\it ( -3273)} (-2885) {\it [-2448, 3720]}  & 1700 & $1\times 10^{-9}$\\
COs1low &center       &23:23:29.8, 58:48:43  &3340006.1, 07-18 & -4690 {\it (--)$^f$}  (--)$^f$ {\it [-1552, 2556]}  &900& $1\times 10^{-10}$ \\
COsipeak & SW shell& 23:23:19.1, 58:47:29    &3340015.1[15.2],  01-18[19] &{\bf-330 {\it (-524)} (216}, 3066) {\it [{\bf 202}, 3206]} & 1200 & $3\times 10^{-10}$\\ 
\enddata
\tablenotetext{a}{Date is month and date of 2009. The observation of 3340005.2 (07-18) had unstable attitude control which 
produced no useful data. 3340007.1 (01-15) towards Si peak produced no significant signal.
The first 5 spectra are high resolution spectroscopy with grism (NG), and the others are with prism (NP).
$^b$Ne ejecta velocity.
$^c$Si at 34.815\,\mic\ velocity components are marked in italic.
Ar and Si show similar velocities because they are at similar ejecta layers. 
$^d$A typical error of temperature is $\pm$300 K.
$^e$Errors on the derived masses range from about 30--40\% for
$T\sim 2,500$~K to a factor of 2--4 for $T\sim 1,000$~K.
$^f$ We marked in bold when the CO
velocities are roughly consistent with those in ejecta,
$^g$The velocity could not be obtained due to lack of emission.}
\label{coproperties}
\end{deluxetable}

The observations took place in January and July 2009 during AKARI
phase 3 (post-helium phase); only the near-infrared channel was
available. The point spread function has a FWHM of 4.7$''$, and the
pixel size is $1.46'' \times 1.46''$. Detailed AKARI AOT configuration
and instrument information is described in the AKARI IRC Data User
Manual for Post-Helium (Phase 3) Mission ver1.1 (Onaka et
al.~2009). We used the grism (NG) or the prism (NP) modes for the
spectroscopy. For the NG observations we used mostly the narrower slit
(Nh) with size $3'' \times 1'$ except for 3340007.1, for which the
wider slit (Ns) of $5'' \times 1'$ was used.  For the NP observations,
we always used the Ns slit.  In the NG mode with the Nh slit, the
spectral resolution is about 0.02\,$\mu$m, which provides a resolving
power of 220 at 4.3\,\mic, while the NG with the Ns slit gives a
resolving power of 150 at 4.3\,\mic. The spectral resolution of the NP
mode varies with the wavelength \citep{ohyama07} 
and is about 40 at 4.5\,\mic\ for the Ns slit.
The AKARI IRC grism spectrum covers
2.4-5\,\mic\ while the prism spectrum extends up to 5.5 $\mu$m.  In
each pointing observation, about 8 exposure cycles were obtained and
the on-source integration time was about 360s. During the
spectroscopic observations, one set of imaging data at 3.2\,\mic\ was
obtained to derive accurate position information. The center position
where the spectrum was extracted is derived from the imaging data.
The summary of the observations is given in Table 1.  Note that the
attitude control of AKARI is not good enough to locate the intended
target position on the Ns or Nh slit. Thus the derived position is
somewhat different from the intended ones.
The indicated position is an average over
the observation and we listed the positions where we extracted 9 sets
of spectra.

The data were processed with the latest version of the spectroscopic
toolkit (version 20101025). We extracted a spectrum from the brightest
part of 6 pixels along the slit (8.76$''$) in each dataset.  Thus the
spectra that we show here are those extracted either from a $8.76''
\times 5''$ (Ns slit) or from a $8.76'' \times 3''$ (Nh slit)
area. Each spectrum was smoothed by 3 pixels both in the spectral and
spatial direction, which does not degrade the spectral resolution.
The relative spectral response is determined to be better than 7\%
around 4.5\,\mic, while the absolute calibration for the Ns and Nh
slit spectroscopy has a larger uncertainty ($\sim$20\%).

We also reanalyzed \spitzer\ IRAC images \citep{ennis06,rho08}. We
subtracted from the IRAC band 2 (4.5\,\mic) image the contribution by
synchrotron emission that we estimated from the IRAC 3.6\,\mic\ map, after
accounting for wavelength- and time-dependent zodiacal emission.
The IRS images of Ar, Ne and Si images are from Rho et
al.~(2008) and Smith et al.~(2009) and the velocities of the ejecta (Table
1) towards the CO positions are estimated based on the results of
\citet{delaney11}.

\section{Results: CO properties}

Fig.~\ref{akarispec} shows the AKARI spectra towards four positions in
Cas A. The spectra clearly show the double-peaked profile
characteristic for the unresolved ro-vibrational band due to the CO
fundamental at 4.65\,\mic.  The CO emission is detected from not only
shocked ejecta but also unshocked ejecta at the center. The {\it
  AKARI} spectra show no other lines than the CO fundamental band
within our spectral resolution. 
We were able to extract a dozen
spectra along the 1$'$ slit length, some of which are combined to
increase the signal-to-noise. The properties of 9 spectra are listed
in Table \ref{coproperties}.

\begin{figure}[!h]
\includegraphics[width=9cm]{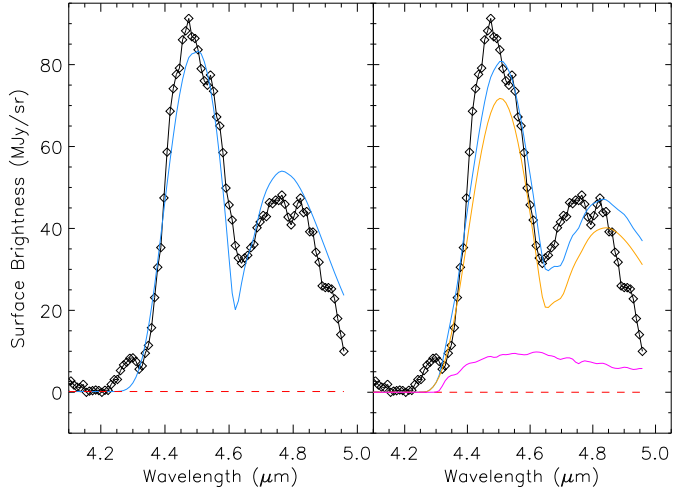}
\caption{High resolution grism spectrum (black)
    towards the southern part (COs1H1 in Table 1). Superposed is the
    best fit CO model (blue) using a single component (left; velocity
    of -4800 km s$^{-1}$) and using two components (orange and purple
    curves) with velocities frozen to those of the Ne ejecta (-1323
    and +1684 km s$^{-1}$). This does not significantly improve the
    quality of the fit, suggesting that the CO may be at a location
    different from the ejecta in this case. }
\label{akarispecmodelfit}
\end{figure}

We have determined the physical parameters of the CO gas by comparing the
observations to isothermal LTE emission models from the SpectraFactory
database\footnote{http://www.spectrafactory.net}\citep{cami10}. The
observed profiles show clearly separated P and R branches, which suggests
that the emission is optically thin (see Fig.~\ref{akarispecmodelfit}). We
therefore present an analysis in the optically thin limit since it greatly
simplifies matters. We also considered optically thick models; however, we
found that all spectra are reproduced more efficiently by optically thin models than
optically thick ones.

In the optically thin limit, the emission is directly proportional to
the number of emitting molecules in the beam, and the observed shape
of the profile (i.e. band width, peak separation and relative strength
of the P and R branches) depends only on the temperature. In LTE, the
Boltzmann equation determines the relative populations over the
molecular levels; and with the frequencies and Einstein A coefficients
given by \citet{goorvitch94} and the partition function from the
HITRAN database \citep[][and references therein]{rothman09},
we then calculate the line intensities for each of the ro-vibrational
transitions that make up the entire band. In the SpectraFactory
models, each of the individual lines is represented by a gaussian line
profile with a Doppler width of 3 \kms. The entire spectrum is then
convolved by a gaussian (instrumental) profile and rebinned to
simulate the observations. While the intrinsic line profile and width
might not be appropriate for the environment studied here, they have
no influence on the resulting band profile. Indeed, the instrumental
resolution is far too low to resolve these individual lines, and since
the lines are optically thin, line overlap is not important.

Finally, we applied a radial velocity shift to the CO model spectra
and included a linear background flux in our model. Our complete
fitting routine then runs over models at temperatures between 500 and
3000 K (in steps of 100 K) and for radial velocities in the range
-10,000 to +10,000 \kms\ (in steps of 10 \kms). For each of these
combinations, we then used a non-negative least-squares routine
\citep[NNLS,][]{1974slsp.book.....L} to find the scale factors for the
CO model and the background that minimizes the $\chi^2$ for each of
the 9 Cas A spectra.

Table 1 summarizes the resulting best fit parameters for all spectra.
We can reproduce the observed emission with optically thin CO emission
at temperatures between 1200 - 2400 K at the shell and 900 K at the
center; an example of the model fits is shown in
Fig.~\ref{akarispecmodelfit}. Such temperature differences make sense:
since the center is where unshocked ejecta (e.g. Si) are detected
\citep{rho08} it is not surprising that we find the cooler CO there.

The derived radial velocities of the CO gas (see Table 1) however
present some puzzles. The different spectra show CO emission at very
different velocities ranging from -300 to -5600 km~s$^{-1}$. Note that
the uncertainty in the NG spectrum is 0.006\,\mic\, which translates
to an uncertainty of typically 400 km~s$^{-1}$ on the velocities. We
find that toward 4 positions (marked in bold in Table 1), the CO
velocities are roughly consistent with those in ejecta, but for the
other positions, there are large differences between the CO velocities
and those measured from the gas lines. CO models where the radial
velocity is fixed to any of the gas line velocities are significantly
worse; a comparison is shown in Fig.~\ref{akarispecmodelfit}. Although
non-LTE effects clearly affect the band shape (see Fig.~\ref{akarispecmodelfit2}),
it may well be that some of the CO velocities intrinsically differ 
from the ionized gas line velocities. 

In the optically thin LTE limit, we can estimate the CO mass as follows.
The total (integrated) flux in the CO band is given by $$F = N {1\over{4
\pi D^2}} \Sigma {g_i~e^{-E_i/kT}A_{ij}~h\nu_{ij}\over{P}} $$ where $P$ is
the partition function, $N$ is the total number of CO molecules in the
beam, $D$ is the distance, $g_i$ is the statistical weight and $E_i$ is the
energy of the upper level $i$, $A_{ij}$ is the Einstein coefficient for
spontaneous transition from $i$ to $j$ and the sum runs over all
transitions in the band. Given the derived CO temperature, $N$ is readily
calculated, and multiplying this by the mass of a CO molecule then yields
the total CO mass per beam.

The total CO mass from all 9 spectra combined is about
1.4$\times$10$^{-8}$ M$_{\odot}$, where the derived mass for each
spectrum is given in Table 1. From the CO fundamental map in
Fig.~\ref{COmap}, we find that the total CO flux is a factor of 43
higher than the total flux of all 9 spectra combined. A rough estimate
of the total CO mass in Cas A is thus (5.9$\pm$0.3)$\times$10$^{-7}$
M$_{\odot}$.

\section{Discussion } 

\begin{figure}
\includegraphics[width=14cm]{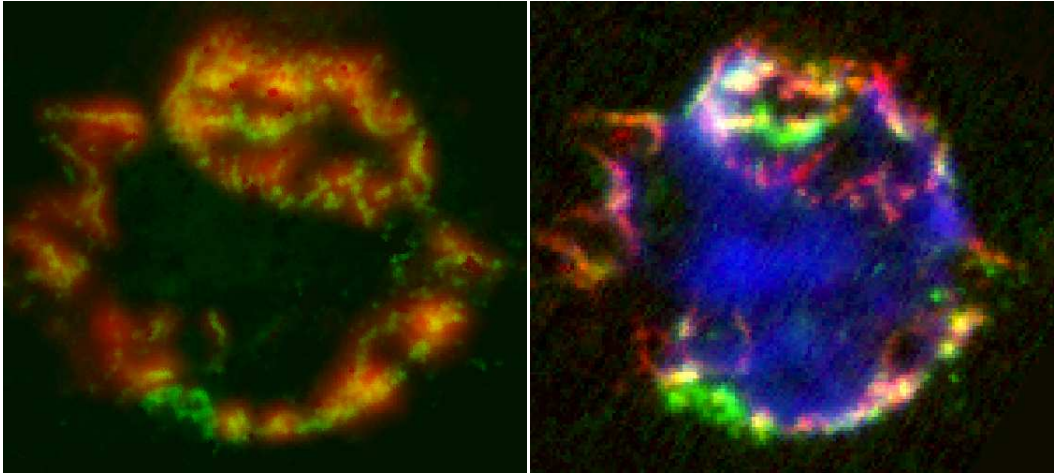}
\caption{
Mosaicked CO and dust maps (a) where
CO molecule map (green) is derived from \spitzer\ IRAC
4.5\,\mic\ map and the 21 \mic\ dust (red, Rho et al.\ 2008).
Mosaicked ejecta maps (b) where red, green and blue represent 
Ar, Ne and Si maps, respectively \citep{rho08, smith09}.
The crescents (excess in green) 
appear in both CO and ejecta maps.
}
\label{COmap}
\end{figure}

Fig.~\ref{COmap} shows the fundamental band CO map.
Surprisingly, the distribution of freshly formed
CO molecules is similar to that of the ejecta, especially when
comparing to the Ne map. This indicates that CO
molecules are formed in ejecta as a precursor of dust formation. The
fact that the morphology of the CO map is almost identical to that of
the ejecta map proves that CO has formed in an earlier phase of the
supernova event, and is now heated by the more recent reverse shock.

We also compare the CO map to the 21 $\mu$m dust map \citep{rho08,
barlow10}. The 21 $\mu$m dust feature can be reproduced mainly with silica
(SiO$_2$) grains with a temperature of $\sim$85 K \citep{rho10} and overall
morphology of the two maps is similar. The CO emission coincides with the
crescents  (see ``green excess" emission in
Fig.~\ref{COmap}) where the ratio of [Ne II] to [Ar II] is high
\citep{smith09}. The 21 $\mu$m dust emission is strongly correlated with Ar
\citep{rho08}, whereas the CO molecules are strongly correlated to Ne
ejecta.
The presence of CO in Cas A implies that less carbon is available to form
carbon dust from the ejecta, because a large portion of carbon may have
been locked in CO. If Type II supernovae such as Cas A are a primary source
of dust in high red-shift galaxies, such dust may be relatively deficient
in carbonaceous grains.

\begin{figure}[!h]
\plotone{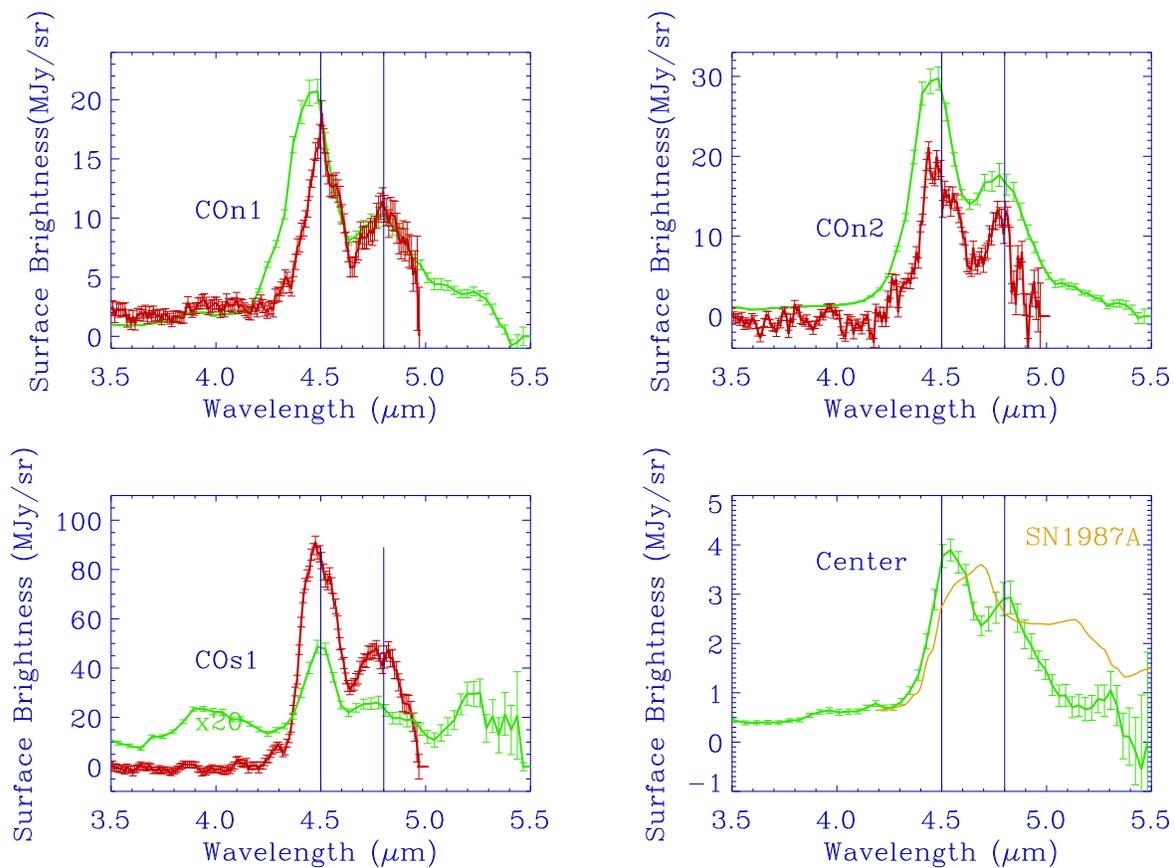}
\caption{
AKARI IRC spectra  of Cas A showing the CO fundamental band detection.
The grism (high resolution) and prism (low) spectra are shown in red and green, respectively.
The coordinates of each position are listed in Table 1.
SN1987A spectrum (orange) from \citet{wooden93} is also shown for comparison.
}
\label{akarispec}
\end{figure}

When comparing the AKARI spectra with those of SN 1987A (see 
Fig.~\ref{akarispec}), there are several notable differences: Cas A lacks
the Pf$\gamma$ and Br$\alpha$ hydrogen lines and the underlying
continuum seen in SN1987A, and there are clear differences in the appearance of the CO features.
In Cas A, the profiles show the double-peak structure that
is characteristic for optically thin CO emission; in SN1987A, on the other
hand, there is no obvious separation between the peaks, indicating that the
CO emission is optically thick \citep[see Fig.~\ref{akarispec}]{liu92}.

\begin{figure}[!h]
\includegraphics[width=7cm]{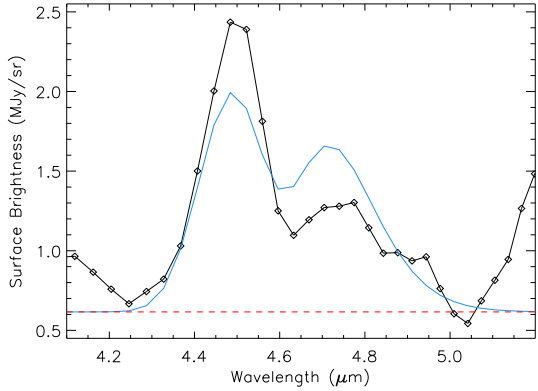}
\caption{Low resolution prism
    spectrum towards the southern part (black; note though that the
    observed position is different from the high resolution spectrum,
    COs1H1) and the best fit CO model. Large discrepancies beween the
    spectrum and the LTE model are indicative of non-LTE effects.}
\label{akarispecmodelfit2}
\end{figure}

For SN1987A, \citet{spyro96} used an LTE model to study the CO
emission in the first overtone (2.2--2.5\,\mic), and estimated a CO
mass of $\rm \sim (2-5) \times 10^{-5} \,M_{\odot}$.  
\citet{liu92} later reanalyzed these data
with a non-LTE model which reproduces the observed CO overtone bands
much better. Consequently, they derived a CO mass of 2.2-2.5$\times
10^{-3} \,M_{\odot}$, 40 - 100 times larger than the LTE estimate.
Also for Cas A, many spectra show non-LTE signatures (see Fig.~\ref{akarispecmodelfit2});
if these would similarly lead to an underestimate of the CO mass by
about the same factor, the CO mass of Cas A could be (2.4-
6)$\times~10^{-5}$ M$_{\odot}$.

In SN environments, several mechanisms are known to
  destroy CO molecules. In the presence of small-scale mixing, it is
  generally believed that CO is destroyed efficiently by collisions
  with helium ions: $He^{+} + CO \rightarrow He + C^+ + O$; if
  He$^{+}$ is absent however, efficient CO removal happens through
  impacts with energetic electrons that can ionize and dissociate CO
  \citep{liu92}. Finally, CO is also destroyed by charge transfer with
  Ne$^+$. In a fully mixed model on the other hand, CO is removed by
  photodissociation and photoionization. When no CO was detected for
  SN1987A at 800 days after the explosion, it was thus concluded that
  all CO had been destroyed \citep{wooden93}. However, the detection
  of CO in Cas A indicates that CO survives or reforms in the ejecta.

The CO detection in Cas A has implications for dust formation in SNe and
the cycle of carbon in the ISM.  First, current dust formation models
\citep{nozawa03} assume that CO molecules are destroyed prior to dust
formation. However, the CO detection shows this is not the case. A
significant amount of carbon may have been locked in CO during the 330-yr
lifetime of Cas A. We have checked ISO LWS observations and found that
[C~II] lines (at 158.7 $\mu$m) toward Cas A do not match the velocities of
the supernova remnant. After subtracting the background, the lack of [C~II]
emission associated with Cas A hints that a large portion of carbon ejecta
is in CO and not ionized carbon.
It awaits confirmation with a higher sensitivity and spectral resolution
such as with Herschel or SOFIA.  Second, the presence of CO in Cas A
suggests that macroscopic mixing in ejecta is very small. Typical electron
densities during the formation of CO are 5$\times$10$^8$ - 10$^6$ cm$^{-3}$
\citep{gearhart99}.  The highly clumpy structures can be explained by such
high density CO gas within the warm or hot ejecta. The HST observations of
Cas A \citep{fesen01} suggest a size between 0.2$''$-0.6$''$ (physical size
of 6000-12000 AU). Assuming that the density of CO is 5$\times$10$^8$ -
10$^{6}$ cm$^{-3}$ and using the estimated CO mass in Table 1 the clump
size would be 2-14 (20-140) AU for a filling factor of 10$^{-1}$
(10$^{-4}$). Note that a CO density could not be directly estimated because
the beam for the spectrum is too large compared to the size of clumps and
thus the uncertainty of the filling factor dominates.


We found CO knots outside the bright ring and near the jet-like
structures, and found that some of the clumps did not have ejecta
counterparts. These dense, molecular ejecta clumps
  propagate into the ISM and will eventually cool down to ISM
  temperatures. Our observations thus provide direct evidence that
  supernovae eject dense clumps into the ISM.

In summary, our observations support the idea that
SNe could be important sites of molecule formation resulting in
species such as CO, SiO, SiS, O$_2$, SO, and CO$_2$
\citep{cherchneff08}. A significant portion of the CO molecules has
survived at least $\sim 330$ years after the initial
explosion. \cite{cherchneff09} suggest that molecules form very
efficiently in the ejecta of supernovae and that 13--34\% of the
progenitor mass of ejecta material could be in molecular form.  For a
25 M$_{\odot}$ progenitor, \cite{woosley02} estimated a gas ejecta
mass of $\sim$15 M$_{\odot}$ after accounting for mass loss; however,
observations of Cas A infer an observed ejecta mass of $\sim$4
M$_{\odot}$ \citep{hwang11}.
If CO locks up C and O in molecular form, and other
  molecular species lock up other heavy elements, then some of the
  missing ejecta mass could be in molecular form as well as in dust
  form. Recently, a dust mass of 0.4-0.7 M$_\odot$ was estimated in SN1987A \citep{matsuura11}, 
emphasizing the importance of dust in supernovae.

We have shown the detection of the fundamental
  vibrational mode of CO at 4.65\,\mic\ from Cas A with AKARI. We
  infer CO temperatures of 900 - 2400 K and a total CO mass of
  (5.9$\pm$0.3)$\times$10$^{-7}$ M$_{\odot}$ using an LTE model. CO
  has not been efficiently destroyed, and molecular species could thus
  be a significant reservoir of heavy elements in the ejecta.  Future
modeling including supernovae explosion models and CO radiative
transport calculations in non-LTE will allow us to examine the core
density profiles and the circumstellar medium profile before the
explosion.

The First Light TEst CAMera (FLITECAM) \citep{mclean06} on SOFIA
covers 1-5$\mu$m, and it will be easier to detect the first overtone
and fundamental CO bands when observing in the stratosphere than from
the ground.  High
resolution spectroscopy in the mid-infrared (EXES) and low-resolution
grism (FORCAST) \citep{adams10} on SOFIA will thus further advance our
understanding of molecule and dust evolution.

\acknowledgements 
We thank Tracey DeLaney for providing the doppler-shifted velocity of ionic
lines. 
This work is based on observations with AKARI, a JAXA project with the
participation of ESA, and made with the  {\it Spitzer Space Telescope}.

{}


\clearpage

\begin{thebibliography}{}

\bibitem[Adams et al.(2010)]{adams10} {Adams}, J.~D., et al.~, 2010, SPIE 7735, 62
\bibitem[Barlow et al(2010)]{barlow10}Barlow, M.J. et al., 2010, A\&A, 518L, 138
\bibitem[Cami et al.(2010)]{cami10}Cami, J., van Malderen, R., \& Markwick, A.J., 2010, ApJS, 187, 409
\bibitem[Cherchneff \& Lilly (2008)]{cherchneff08}  
 Cherchneff, I. \& Lilly, S., 2008, ApJ, 683, L123  
\bibitem[Cherchneff \& Dwek (2009)]{cherchneff09}
 Cherchneff, I. \& Dwek, I., 2009, ApJ, 703, 642 
Meyer, B. S., 2001, \apj, 562, 480
\bibitem[DeLaney et al.(2011)]{delaney11} DeLaney, T. A. et al., 2011, ApJ, 725, 2038
\bibitem[Ennis et al.(2006)]{ennis06}Ennis, J., Rudnick, L., Reach, W.T., Smith, J.D., Rho, J., DeLaney, T.D., Gomez, H., Kozasa, T., 2006, ApJ, 652, 376
\bibitem[Fesen(2001)]{fesen01}Fesen, R. A., 2001, ApJS, 133, 161 
\bibitem[Gearhart et al.(1999)]{gearhart99}Gearhart, R. A., Wheeler, J. C., Swartz, D. A., 1999, \apj, 510, 944
\bibitem[Gerardy et al.(2002)]{gerardy02} 
Gerardy, C. L., Fesen, R. A.,  Nomoto, K., Meada, K., Hoflich, P., \& Wheeler, J.C., 2002, PASJ, 54, 905 
\bibitem[Goorvitch(1994)]{goorvitch94} Goorvitch, D. 1994, ApJS, 95, 535
\bibitem[Hwang et al.(2011)]{hwang11} Hwang, U. \& Laming, J. M., 2011, ApJ, in press  
\bibitem[Kotak et al.(2006)]{kotak06}Kotak, R. et al. 2006, ApJ, 651, L117  
\bibitem[{Lawson} \& {Hanson}(1974){Lawson} \& {Hanson}]{1974slsp.book.....L}
{Lawson} C.L., {Hanson} R.J., 1974, {Solving least squares problems}, 
  Prentice-Hall Series in Automatic Computation, Englewood Cliffs:
  Prentice-Hall, 
\bibitem[Liu, Dalgarno \& Lepp(1992)]{liu92} Liu, W., Dalgarno, A., \& Lepp, S., 1992, \apj, 396, L679 
\bibitem[Liu \& Dalgarno(1995)]{liu95} Liu, W., \& Dalgarno, A., 1995, \apj, 454, L472
\bibitem[Matsuura et al.(2011)]{matsuura11}
Matsuura, M. et al., 2011, Science, 333, 1258
\bibitem[McLean et al.(2006)]{mclean06} {McLean}, I.~S., et al., 2006, SPIE, 6269, 168
\bibitem[Nozawa et al.(2003)]{nozawa03}Nozawa, T., Kozasa, T., 
Umeda, H., Maeda, K., \& Nomoto, K., 2003, ApJ, 598, 785
\bibitem[Ohyama et al.(2007)]{ohyama07}Ohyama, Y. et al. 2007, \pasj, 59, S411
\bibitem[Onaka et al.(2007)]{onaka07}Onaka, T. et al. 2007, \pasj, 59, S401

\bibitem[Rho et al.(2008)]{rho08}Rho, J., Kozasa, T., Reach, W.T., Smith, J., Rudnick, L., DeLaney, T., Ennis, J., Gomez, H., Tappe, A., 2008, ApJ, 673, 271 
\bibitem[Rho et al.(2009)]{rho09}
Rho, J., Jarrett, T. H., Reach, W. T., Gomez, H., \& Andersen, M., 2009, ApJL, 693, L39
\bibitem[Rho et al.(2010)]{rho10}
Rho, J., Gomez, H., Lagage, P.-O., Boogert, A., Reach, W. T., \& Dowell, D., 2010,
38th COSPAR Scientific Assembly. July 2010, in Bremen, Germany, E19-0105-10
\bibitem[Rothman et al.(2009)]{rothman09}
Rothman, L., et al. 2009, J. Quant. Spectrosc. Radiat. Transfer, 110, 533
\bibitem[Smith et al.(2009)]{smith09} 
Smith, J. D. et al.~2009, ApJ, 693, 713
Isensee, K.A 2009, ApJ, 693, 713
\bibitem[Spyromilio et al.(1988)]{spyromilio88}Spyromilio, J.,  
Meikle, W. P. S., Learner, R.C.M., \& Allen, D.A., 1988, Nature, 334, 327
\bibitem[Spyromilio \& Leibundgut(1996)]{spyro96}Spyromilio, J., Leibundgut, B., 1996, MNRAS, 283, L89
\bibitem[Wooden et al.(1993)]{wooden93}Wooden, D.H. et al., 1993, ApJS, 88, 477
\bibitem[Woosley, Heger, \& Weaver (2002)]{woosley02} Woosley, S. E., Heger, A., \& T. A. Weaver, 2001, Reviews of M. Physics, 74, 1015.

\end{thebibliography}
\end{document}